\documentclass[12pt]{article}
\usepackage{amssymb}
\usepackage{graphicx,psfrag}
\usepackage{color}
\usepackage[nosort]{cite}

\newcommand \bra[1]{\left< {#1} \,\right\vert}
\newcommand \ket[1]{\left\vert\, {#1} \, \right>}

\newcommand{\bea}{\begin{eqnarray}}
\newcommand{\eea}{\end{eqnarray}}
\newcommand{\simgt}{\hbox{ \raise3pt\hbox to 0pt{$>$}\raise-3pt\hbox{$\sim$} }}
\newcommand{\simlt}{\hbox{ \raise3pt\hbox to 0pt{$<$}\raise-3pt\hbox{$\sim$} }}

\newcommand{\clfn}{\setcounter{footnote}{0}}
\newcommand{\LQ}{\Lambda_{\rm QCD}}

\definecolor{black}{rgb}{0,0,0}
\definecolor{red}{rgb}{.8,0,0}
\definecolor{magenda}{rgb}{1,0,1}
\definecolor{blue}{rgb}{0,0,1}
\definecolor{brown}{cmyk}{0,0.81,1,0.70}
\definecolor{green}{rgb}{0,0.8,0}

\def\red{\color{red}}

\def\blue{\color{blue}}

\begin{document}

\begin{titlepage}
\title{\Large
\vspace{28mm}
Perturbative QCD Potential
and String Tension\thanks{
Talk given at the ``13th International Seminar on High Energy Physics
(Quarks 2004),''  Pushkinskie Gory, Russia, May 24 -- 30, 2004.
}\vspace{7mm}}
\author{
Y.~Sumino
\\ \\ \\ Department of Physics, Tohoku University\\
Sendai, 980-8578 Japan
}
\date{}
\maketitle
\thispagestyle{empty}
\vspace{-4.5truein}
\begin{flushright}
{\bf TU--732}\\
{\bf October 2004}
\end{flushright}
\vspace{4.5truein}
\begin{abstract}
\noindent
{\small
The leading non-perturbative contribution
to the static QCD potential at $r \ll \LQ^{-1} $ is known to be ${\cal O}(r^2)$
in operator-product expansion.
It indicates that a ``Coulomb+linear" potential
at $r \simlt \LQ^{-1}$
is included in the perturbative QCD prediction of the potential.
It was shown that this is indeed the case, and
the ``Coulomb+linear" potential has been systematically computed
up to next-to-next-to-leading logarithmic order, 
in terms of the Lambda parameter in the $\overline{\rm MS}$ scheme 
($\Lambda_{\overline{\rm MS}}$).
We review the present status of the perturbative prediction
for the QCD potential, which takes into account the
theoretical breakthrough that took place around 1998.
}
\end{abstract}
\vfil

\end{titlepage}


\section{Introduction}
In this article we review the present status of perturbative QCD predictions 
for the QCD potential,
in the distance region relevant to the bottomonium and charmonium
states, 
$0.5~{\rm GeV}^{-1} (0.1~{\rm fm}) \simlt r \simlt
5~{\rm GeV}^{-1} (1~{\rm fm})$.
We take into account the theoretical breakthrough that took place 
around 1998 \cite{renormalon},
which improved the accuracy of perturbative prediction dramatically.

For a quite long time, the perturbative QCD predictions of 
the QCD potential $V_{\rm QCD}(r)$
were {\it not} successful 
in the above distance region.
In fact, the perturbative series turned out to be very poorly convergent at
$r \simgt 0.5~{\rm GeV}^{-1}$; 
uncertainty of the series is so large that one could hardly obtain
meaningful predictions.
Even if one tries to improve the perturbation series by
certain resummation prescription (such as renormalization
group improvement),
scheme dependence of the result was also very large \cite{grunberg};
hence, one could neither
obtain accurate prediction of the potential in this distance region.
It was later pointed out that the large uncertainty of the perturbative 
QCD prediction can be understood as caused by
the leading-order [${\cal O}(\LQ)$] infrared (IR) renormalon contained
in $V_{\rm QCD}(r)$ \cite{al}.

\begin{figure}[t]
\begin{center}
\psfrag{r}{\raise1pt\hbox{$r~~[{\rm GeV}^{-1}]$}}
\psfrag{Potential}{\hspace{-3mm} Potential~~~~[GeV]}
\psfrag{Coulomb Pot.}{Coulomb Pot.}
\psfrag{Phenomenological Pot.}{\blue{Phenomenological Pot.}}
\psfrag{breakdown}{\red{\small breakdown of pert.\ prediction}}
\includegraphics[width=7.5cm]{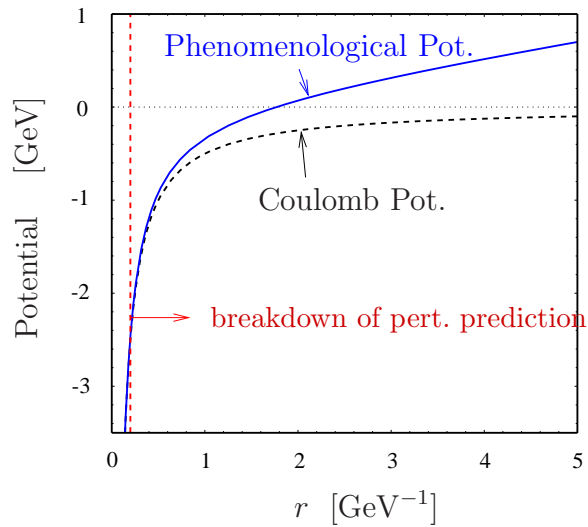}
\label{Vc-Vpheno}
\caption{Status of theoretical predictions for the QCD potential before 
the discovery of renormalon cancellation.
One could hardly see sign of convergence of perturbative series
at $r \simgt 0.5~{\rm GeV}^{-1}$.
}
\end{center}
\end{figure}

On the other hand,
empirically it has been known that phenomenological potentials
and lattice computations of $V_{\rm QCD}(r)$ are both
approximated well by the sum of a Coulomb potential and a linear
potential in the above range $0.5~{\rm GeV}^{-1} \simlt r \simlt
5~{\rm GeV}^{-1}$.  (See e.g.\ \cite{bali}).
The linear behavior of $V_{\rm QCD}(r)$ at large distances
$r \gg \LQ^{-1}$
(verified numerically by lattice computations)
is consistent with the quark confinement picture.
For this reason,  and given the very poor predictability of perturbative
QCD,
it was often said that, while the ``Coulomb'' part of $V_{\rm QCD}(r)$  
(with logarithmic correction at short-distances) is
contained in the perturbative QCD prediction, the linear part
is purely non-perturbative and absent in
the perturbative QCD prediction even at $r \simlt \LQ^{-1}$, and that
the linear potential needs to be added to the perturbative prediciton to obtain
the full QCD potential.
See Fig.~\ref{Vc-Vpheno}.
Nevertheless, to the best of our knowledge, there was no firm theoretical basis
for this argument.

Several years ago, the perturbative prediction of the QCD potential
became much more accurate in this region.
There were two important developments:
(1) The complete ${\cal O}(\alpha_S^3)$ corrections to the
QCD potential have been computed \cite{ps};
also the relation between the quark pole mass and the $\overline{\rm MS}$
mass has been computed up to ${\cal O}(\alpha_S^3)$ \cite{mr}.
(2) A renormalon cancellation was discovered
\cite{renormalon} in 
the total energy of a static 
$Q\bar{Q}$ pair\footnote{
By ``static'', we mean that the kinetic energies of $Q$ and $\bar{Q}$
are neglected.
},
$E_{\rm tot}(r) \equiv 2 m_{\rm pole} + V_{\rm QCD}(r)$.
Consequently, convergence of the
perturbative series of $E_{\rm tot}(r)$
improves drastically,
if it is expressed in terms of the quark $\overline{\rm MS}$ mass
instead of the pole mass.

Physically, improvement of convergence stems from decoupling of IR gluons
from the color-singlet $Q\bar{Q}$ system.
Intuitively,
IR gluons, whose wavelengths are larger than $r$, cannot see the color charge
of $Q$ or $\bar{Q}$ but only see the total charge of this system.
Therefore, in the computation of the total energy of the 
$Q\bar{Q}$ system, we expect that contributions of IR gluons should decouple.
This is naturally realized, if we use a quark mass, which is renormalized
to include contributions
of only UV gluons, such as the $\overline{\rm MS}$ mass;
in this case, there is a cancellation of IR contributions 
between the self-energies of $Q/\bar{Q}$ and the potential
energy.
If we use a mass, which includes also contributions of IR gluons,
such as the pole mass,\footnote{
Pole mass is defined as the energy of a free quark at rest.
Hence,  IR gluons can see its color charge and contribute to the pole mass.
} 
then the decoupling of IR gluons is not realized.
Generally, convergence of a perturbative expansion is worse
when there are more contributions from IR gluons, and vice versa.

\begin{figure}[t]\centering
\psfrag{mu}{\hspace{0mm}$\mu$~[GeV]}
\psfrag{Etot}{\hspace{2mm}$E_{\rm tot}(r)$~~[GeV]}
\psfrag{r = 2.5}{\hspace{0mm}$r=2.5$~GeV$^{-1}$}
\psfrag{Pole-mass scheme}{\hspace{0mm}\raise-0pt\hbox{Pole-mass scheme}}
\psfrag{MSbar-mass scheme}{\hspace{0mm}$\overline{\rm MS}$-mass scheme}
\includegraphics[width=6cm]{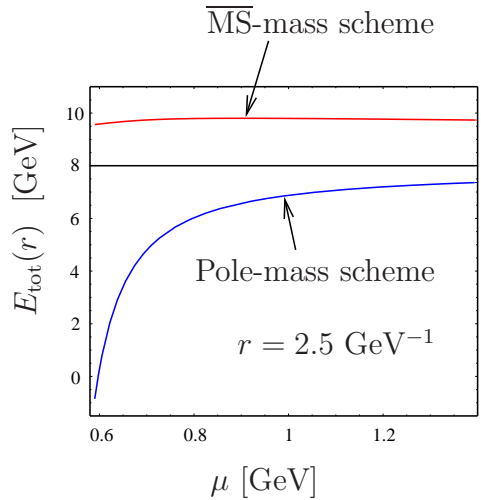}
\caption{\small 
Scale dependences of $E_{{\rm tot}}^{b\bar{b}}(r)$ 
up to ${\cal O}(\alpha_S^3)$ at 
$r =2.5~{\rm GeV}^{-1}$, in the pole-mass and 
$\overline{\rm MS}$-mass schemes.
A horizontal line at $8$~GeV is plotted as a guide.
\label{scale-dep}}
\end{figure}

Let us demonstrate the improvement of accuracy of the perturbative
prediction for the total energy
$E_{\rm tot}(r) $ 
up to ${\cal O}(\alpha_S^3)$, when the cancellation of
${\cal O}(\LQ)$ renormalons is incorporated.
As an example, we take the bottomonium case:
We choose the $\overline{\rm MS}$
mass of the $b$-quark, renormalized at the  $b$-quark $\overline{\rm MS}$
mass, as
$\overline{m}_b \equiv 
m_b^{\overline{\rm MS}}(m_b^{\overline{\rm MS}})
= 4.190$~GeV;
in internal loops, four flavors of light quarks are included
with $\overline{m}_u=\overline{m}_d=\overline{m}_s=0$
and $\overline{m}_c=1.243$~GeV.
(See the formula for $E_{{\rm tot}}^{b\bar{b}}(r)$ in \cite{Recksiegel:2001xq}.)
In Fig.~\ref{scale-dep}, we fix $r=2.5~{\rm GeV}^{-1}$
(midst in the distance range of our interest)
and examine the renormalization-scale ($\mu$) dependences of $E_{{\rm tot}}(r)$.
We see that $E_{{\rm tot}}(r)$ is much less scale dependent when we
use the $\overline{\rm MS}$ mass  than when we use the pole mass.
Hence, the perturbative prediction of $E_{{\rm tot}}(r)$
is much more
stable in the former scheme.

We also compare the convergence behaviors
of the perturbative series of $E_{\rm tot}$ for the same $r$
and when $\mu$ is fixed to the value, at which $E_{\rm tot}$
becomes least sensitive to variation of $\mu$ 
(minimal-sensitivity scale).
Convergence of the perturbation series turns out to be close to optimal
for this scale choice:
at $r =2.5~{\rm GeV}^{-1}$, the minimal-sensitivity scale is
$\mu = 0.90$~GeV.
We find
\bea
E_{{\rm tot}}^{b\bar{b}}(r) &=& 
10.408 - 0.275 - 0.362 - 0.784 ~~{\rm GeV}
~~
\mbox{(Pole-mass scheme)}
\\
&=&
~\,8.380 + 1.560 -0.116 - 0.022~~{\rm GeV}
~~
\mbox{($\overline{\rm MS}$-mass scheme)} .
\eea
The four numbers represent the ${\cal O}(\alpha_S^0)$,
${\cal O}(\alpha_S^1)$, ${\cal O}(\alpha_S^2)$ and 
${\cal O}(\alpha_S^3)$ terms of the series expansion in each
scheme.
The ${\cal O}(\alpha_S^0)$ terms represent merely the twice of
the pole mass and of the $\overline{\rm MS}$ mass, respectively.
As can be seen, if we use the pole mass, the series is not
converging beyond ${\cal O}(\alpha_S^1)$.
On the other hand,
in the $\overline{\rm MS}$-mass scheme, the series is converging.
One may also verify that, 
when the series is converging ($\overline{\rm MS}$-mass scheme),
$\mu$-dependence of $E_{\rm tot}(r)$ decreases
as we include more terms of the perturbative series,
whereas when the series is diverging
(pole-mass scheme), $\mu$-dependence does not decrease with
increasing order.
(See e.g.\ \cite{topmass}.)

We observe qualitatively the same features at
different $r$ and for different number of light quark
flavors $n_l$, or when we change values 
of the masses $\overline{m}_b$, $\overline{m}_c$.
Generally, at smaller $r$,
$E_{\rm tot}(r)$ becomes less
$\mu$-dependent and more convergent, 
due to asymptotic freedom.
See \cite{Recksiegel:2001xq,Recksiegel:2002um} for details.

The aim of this paper is to review the properties of $E_{\rm tot}(r)$,
given the much more accurate prediction as compared to several
years ago.
In Sec.~2 we examine $E_{\rm tot}(r)$ up to ${\cal O}(\alpha_S^3)$.
Sec.~3 provides a classification of perturbative and non-perturbative
contributions to $E_{\rm tot}(r)$ in terms of operator-product-expansion (OPE).
We present our main result, that the perturbative prediction of the
leading short-distance contribution to $E_{\rm tot}(r)$ is given as
a ``Coulomb+linear'' potential, in Sec.~4.
Conclusions are given in Sec.~5.

\section{\boldmath Pert.\ prediction of $E_{\rm tot}(r)$ up to ${\cal O}(\alpha_S^3)$}
\clfn

Let us first review comparisons of the 
perturbative predictions of $E_{\rm tot}(r)$ up to ${\cal O}(\alpha_S^3)$
with phenomenological potentials and with lattice computations
of the QCD potential.

\begin{figure}[t]\centering
\psfrag{XXX}{\footnotesize $r~[{\rm GeV}^{-1}]$}
\psfrag{YYY}{\footnotesize \hspace{-8mm} Energy~~~$[{\rm GeV}]$}
\psfrag{PowerLaw}{\footnotesize Power--law potential}
\psfrag{Log}{\footnotesize Log potential}
\psfrag{Cornell}{\footnotesize Cornell potential}
\psfrag{1161}{\footnotesize $\alpha_s(M_Z)=0.1161$}
\psfrag{1181}{\footnotesize $\alpha_s(M_Z)=0.1181$}
\psfrag{1201}{\footnotesize $\alpha_s(M_Z)=0.1201$}
\psfrag{u1s}{\footnotesize $\Upsilon(1S)$}
\psfrag{u2s}{\footnotesize $\Upsilon(2S)$}
\psfrag{u3s}{\footnotesize $\Upsilon(3S)$}
\psfrag{u4s}{\footnotesize $\Upsilon(4S)$}
\psfrag{j1s}{\footnotesize $J/\psi$}
\psfrag{j2s}{\footnotesize $\psi(2S)$}
\includegraphics[width=90mm]{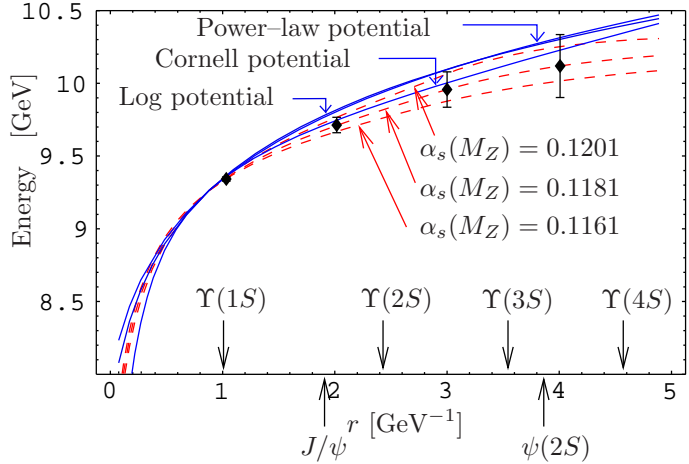}
\caption{\small $E_{\rm tot}(r)$ and typical
phenomenological potentials.
Constants are added to make all curves coincide
at $r=1~{\rm GeV}^{-1}$.
Arrows at the bottom show the r.m.s.\ radii of the heavy
quarkonium states.
The figure is taken from \cite{Recksiegel:2001xq}.
}
\label{comparemodels}
\end{figure}

In Fig.~\ref{comparemodels}, $E_{\rm tot}(r)$ is compared with typical
phenomenological potentials.
Since the latter are determined from
the heavy quarkonium spectra, realistic values for
the input parameters of $E_{\rm tot}(r)$,
as given in the previous section, have been chosen.
The scale $\mu =\mu(r)$ is fixed by either
$\partial E_{\rm tot}/\partial \mu=0$ 
(minimum sensitivity scale) or
$|E_{\rm tot}^{(3)}|=\mbox{minimum}$, but both
prescriptions lead to almost same values of $E_{\rm tot}(r)$.
See \cite{Sumino:2001eh,Recksiegel:2001xq,Recksiegel:2002um} for details.
We see that $E_{\rm tot}(r)$ corresponding to the present
values of the strong coupling constant (dashed lines)
agree well with the phenomenological potentials
within estimated perturbative uncertainties
(indicated by error bars).
We also note that the agreement is lost quickly if we take
$\alpha_S(M_Z)$ outside of the present world-average
values, so that the agreement is unlikely to be accidental.

In fact, by now several works have confirmed agreements 
of the perturbative predictions of the QCD potential
with phenomenological potentials or with lattice computations
\cite{Sumino:2001eh,Necco:2001gh,Recksiegel:2001xq,Pineda:2002se,Lee:2002sn,Recksiegel:2002um}.
Although details depend on how the renormalon
in the QCD potential is cancelled, qualitatively the same conclusions
were drawn: i.e.\ perturbative predictions become accurate
and agree with phenomenological potentials/lattice results
up to much larger $r$ than before.
In particular, in the differences 
between the perturbative predictions
and phenomenological potentials/lattice results,
a linear potential of
order $\Lambda_{\rm QCD}^2 r$ 
at distances $r \simlt \Lambda_{\rm QCD}^{-1}$
was ruled out numerically.
In other words, one cannot get the full QCD potential if one adds a
linear potential to the perturbative QCD potential (after renormalon cancellation).

A crucial point is that,
once the ${\cal O}(\Lambda_{\rm QCD})$ renormalon is cancelled and
the perturbative prediction is made accurate,
the perturbative potential becomes steeper than
the Coulomb potential as $r$ increases.
This feature is understood, within perturbative QCD, 
as an effect of the {\it running} of the strong coupling constant 
\cite{Brambilla:2001fw,Sumino:2001eh,Necco:2001gh}.
In short, the interquark force becomes stronger than the Coulomb force
at larger $r$ by this effect.

\section{Operator-product expansion of \boldmath $V_{\rm QCD}(r)$}

Operator-product expansion (OPE) 
of $V_{\rm QCD}(r)$ for $r \ll \Lambda_{\rm QCD}^{-1}$ 
was developed \cite{Brambilla:1999qa} within an 
effective field theory  
``potential non-relativistic QCD'' (pNRQCD)
\cite{Pineda:1997bj}.
In this framework, $V_{\rm QCD}(r)$ is expanded in $r$
(multipole expansion),
when the following hierarchy of scales exists:
\bea
\Lambda_{\rm QCD} \ll \mu_f \ll \frac{1}{r} .
\label{hierarchy}
\eea
Here, $\mu_f$ denotes the factorization scale.
At each order of the expansion in $r$, short-distance contributions 
($q>\mu_f$) are factorized
into perturbatively computable Wilson coefficients 
and long-distance contributions ($q<\mu_f$) into matrix elements of operators,
i.e.\ non-perturbative quantities.
The leading non-perturbative contribution to the potential turns out
to be ${\cal O}(\Lambda_{\rm QCD}^3 r^2)$.

Explicitly, the QCD potential
is given by
\bea
&&
V_{\rm QCD}(r) = V_S(r;\mu_f) + \delta E_{\rm US}(r;\mu_f),
\label{OPE}
\\ &&
\delta E_{\rm US}=
- i g_S^2 \frac{T_F}{N_C}
\int_0^\infty \! \! dt \, e^{-i \, \Delta V(r)\, t} \,
\bra{0} \vec{r}\!\cdot\!\vec{E}^a(t) \, \varphi_{\rm adj}(t,0)^{ab}\,
\vec{r}\!\cdot\!\vec{E}^b(0) \ket{0}
\nonumber\\&&
~~~~~~~~~~+~
{\cal O}(r^3) .
\label{deltaEUS}
\eea
The leading short-distance contribution to $V_{\rm QCD}(r)$
is given by the singlet potential
$V_S(r)$.
It is a Wilson coefficient, which represents 
the potential between the static $Q\bar{Q}$ pair in
color singlet state.
The leading non-perturbative contribution
[${\cal O}(r^2)$ in the multipole expansion] is contained in the matrix element in
eq.~(\ref{deltaEUS}).
$\Delta V(r) = V_O(r) - V_S(r)$ denotes the
difference between the octet and singlet potentials;
see \cite{Brambilla:1999qa} for details.

The singlet potential $V_S(r;\mu_f)$ can be computed in
perturbative expansion in $\alpha_S$ by matching pNRQCD to QCD.
It is essentially the perturbative expansion
of $V_{\rm QCD}(r)$ after the contribution of soft degrees of freedom
($q<\mu_f$)  is subtracted.
In particular, the IR divergences and IR renormalons contained in
the perturbative expansion of $V_{\rm QCD}(r)$ have been subtracted
via appropriate renormalization prescription  \cite{Brambilla:1999qa,Pineda:2002se,ope}.
Thus, $V_S(r;\mu_f)$ can be computed
accurately by perturbative QCD.

We may understand why the leading non-perturbative contribution is ${\cal O}(r^2)$
as follows.
As well known,
the leading interaction (in expansion of $r$) between soft gluons and a color-singlet 
$Q\bar{Q}$ state
of size $r$ is given by the dipole interaction
$\vec{r}\cdot \vec{E}^a$, where $\vec{E}^a$ denotes the color electric
field.
It turns the color singlet $Q\bar{Q}$ state into a color octet $Q\bar{Q}$ state
by emission of soft gluon(s).
To return to the color singlet $Q\bar{Q}$ state, it needs to reabsorb the
soft gluon(s), which requires an additional dipole interaction.
Thus, the leading contribution of soft gluons to the 
total energy is ${\cal O}(r^2)$.

\section{Pert.\ prediction of \boldmath $V_S(r)$ up to NNLL:\\
``Coulomb+linear" potential}
\clfn

\begin{figure}
\begin{center}
\begin{tabular}{cc}
\psfrag{C1}{$C_1$} 
\psfrag{q}{$q$} 
\psfrag{q*}{$q_*$} 
\psfrag{0}{\hspace{-1mm}\raise-1mm\hbox{$0$}}
\includegraphics[width=4.5cm]{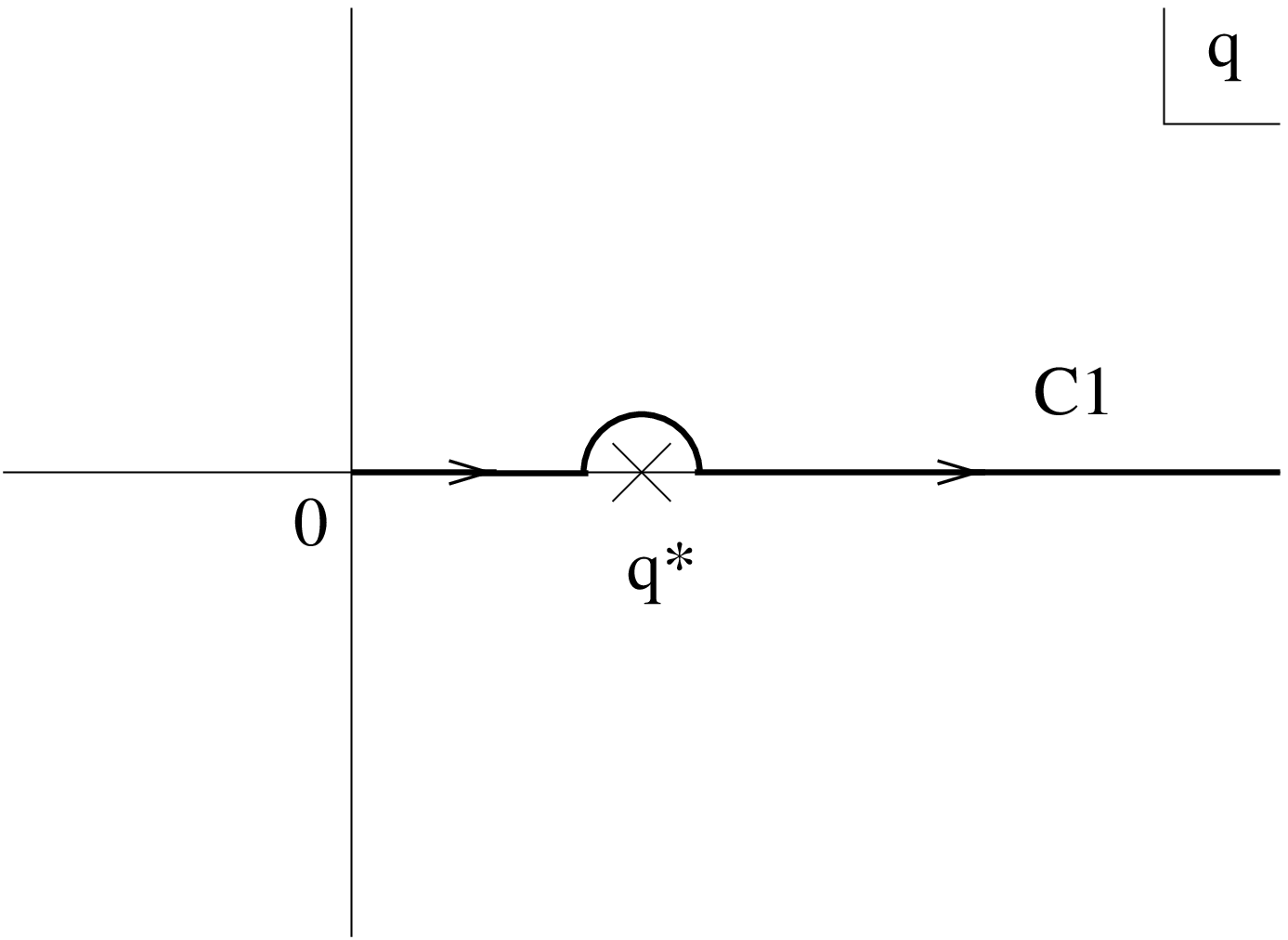} 
& \hspace{20mm}
\psfrag{C2}{$C_2$} 
\psfrag{q}{$q$} 
\psfrag{q*}{$q_*$} 
\psfrag{0}{\hspace{-1mm}\raise-1mm\hbox{$0$}}
\includegraphics[width=4.5cm]{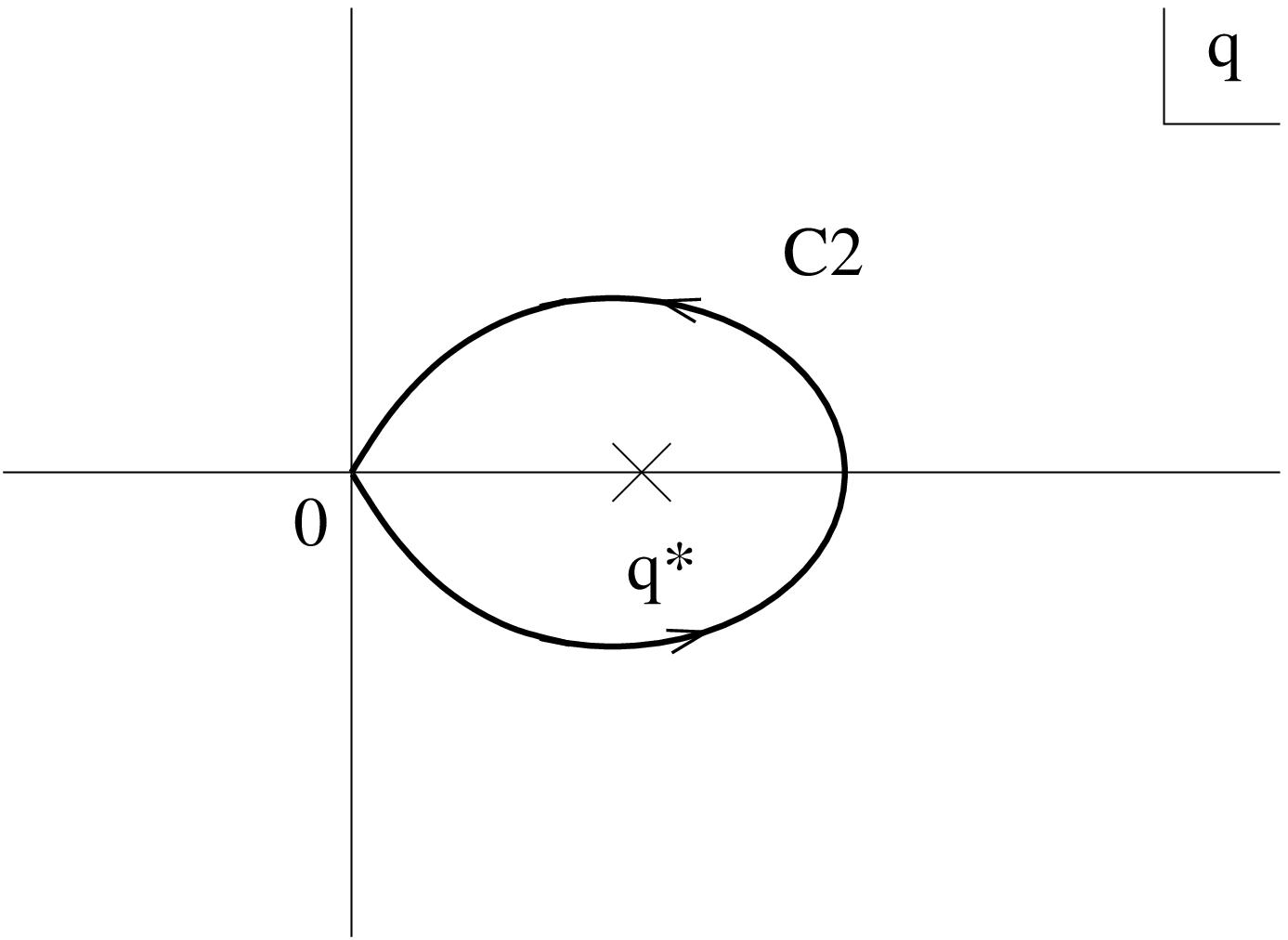}
\end{tabular}
\end{center}
\vspace*{-.5cm}
\caption{\small Integral contours $C_1$ and $C_2$ on the complex $q$-plane.
$q_*$ denotes the Landau singularity of $\alpha_S(q)$.
For 1-loop running, $q_*$ is a pole; for 2- and 3-loop running,
$q_*$ is a branch point.
In the latter case, branch cut is on the real axis starting from $q_*$
to $-\infty$.
\label{path}}
\end{figure}

Here, we present our main result.
As already stated,
the singlet potential $V_{S}(r;\mu_f)$ can be computed in perturbative QCD, 
which may be improved via renormalization group (RG).
We have shown that $V_{S}(r;\mu_f)$ thus computed
can be expressed as a ``Coulomb+linear'' potential,
up to an ${\cal O}(r^2)$ correction.
The correction can be absorbed into
the ${\cal O}(r^2)$ non-perturbative contribution $\delta E_{\rm US}(r;\mu_f)$
by appropriate renormalization prescription.
Explicitly, we find \cite{Sumino:2003yp,ope}
\bea
&&
V_{S}(r;\mu_f) = {\rm const.}+V_{\rm C}(r) + \sigma \, r + {\cal O}(\mu_f^3r^2), 
\label{CplusLpot}
\eea
where
\bea
&&
V_{\rm C}(r) = - \frac{4\pi C_F}{\beta_0 r} 
- \frac{2C_F}{\pi} \, {\rm Im}
\int_{C_1}\! dq \, \frac{e^{iqr}}{qr} \, \alpha_{V}(q) 
,
\label{Vc}
\\ &&
{\sigma} = \frac{C_F}{2\pi i} \int_{C_2}\! dq \, q \, \alpha_{V}(q) .
\label{sigma}
\eea
In the above equations, $\alpha_{V}(q)$ denotes the perturbative evaluation
of the $V$-scheme coupling constant in momentum space \cite{ps} 
(improved by RG,
after subtraction of IR divergences via renormalization).
The integral paths $C_1$ and $C_2$ are displayed 
in Figs.~\ref{path}.
$V_{\rm C}(r)$ and $\sigma$ are defined to be independent of $\mu_f$ (at least) up to
NNLL.
$C_F=4/3$ denotes a color factor;
$\beta_0=11-2n_l/3$ represents the 1-loop coefficient of the
beta function.
In eq.~(\ref{CplusLpot}),
we are not concerned about 
the constant ($r$-independent) part of $V_{S}(r)$, keeping in mind
that it can always be absorbed into $2m_{\rm pole}$ in the
total energy $E_{\rm tot}(r)$.

By evaluating the above integral,
the coefficient of the linear potential $\sigma$ 
can be expressed
analytically in terms of the Lambda parameter in the 
$\overline{\rm MS}$-scheme
$\Lambda_{\overline{\rm MS}}$.
For instance, at LL, it is given by
\bea
\sigma_{\rm LL} = \frac{2\pi C_F}{\beta_0} \,
\left( \Lambda_{\overline{\rm MS}}^{\rm 1-loop} \right)^2 .
\eea

The ``Coulomb'' potential has a short-distance
asymptotic behavior consistent with RG,
\bea
V_{\rm C}(r) \sim -\frac{2\pi C_F}{\beta_0r
\ln\Bigl(\frac{1}{r \Lambda_{\overline{\rm MS}}}\Bigr)} 
~~~~~
(r \to 0) ,
\eea
whereas its long-distance behavior is given by
\bea
V_{\rm C}(r) \sim - \frac{4\pi C_F}{\beta_0r} 
~~~~~
(r \to \infty) .
\eea
In the intermediate region both asymptotic forms are smoothly
interpolated.

\begin{figure}[t]\centering
  \hspace*{-2.5cm}
  \includegraphics[width=14cm]{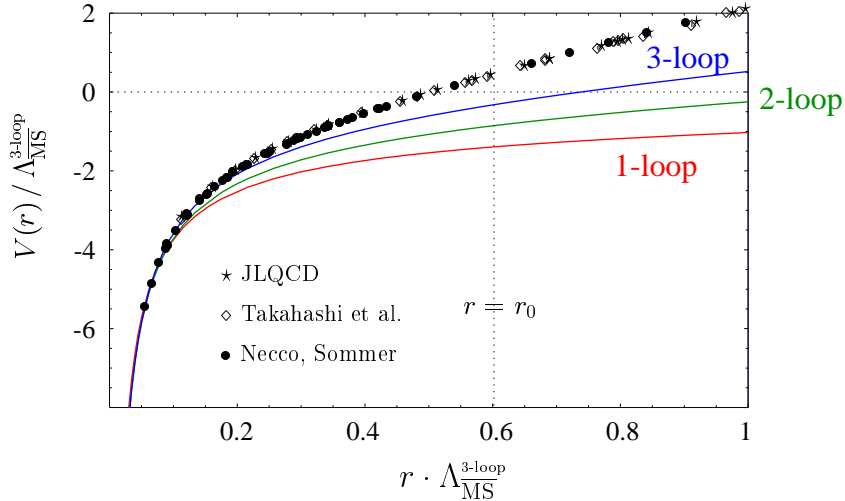}
\caption{\footnotesize
Comparison of the Coulomb+linear potential ($n_l=0$)
with the lattice results in the quenched approximation.
1-, 2- and 3-loop correspond, respectively, to 
LL, NLL and NNLL in the RG-improved computation
of $V_S(r)$.
      \label{comp-lat}}
\end{figure}

Fig.~\ref{comp-lat} shows the Coulomb+linear potentials
corresponding to the RG improvement by
the 1-loop running (LL), 2-loop running
with 1-loop non-logarithmic term (NLL), and 3-loop running
with 2-loop non-logarithmic term (NNLL).
They are compared with the lattice results.
Since the 3-loop non-logarithmic term is not yet known,
the 3-loop running case represents our current best knowledge.
Up to this order, the Coulomb+linear
potential agrees with the lattice results up to larger $r$
as we increase the order.\footnote{
The NNLL originating from the ultra-soft scale \cite{USLL} hardly
changes the 3-loop running case displayed in Fig.~\ref{comp-lat}
\cite{Sumino:2003yp}.
}

\section{Conclusions}

After discovery of renormalon cancellation, perturbative prediction of 
the total energy
$E_{\rm tot}(r)=2m_{\rm pole} + V_{\rm QCD}(r)$ became much more accurate
in the distance range $0.5~{\rm GeV}^{-1} \simlt
r \simlt 5~{\rm GeV}^{-1}$.
Consequently we observe the following:
\begin{itemize}
\item
Perturbative prediction of
$E_{\rm tot}(r)$ up to ${\cal O}(\alpha_S^3)$ agrees well with phenomenological
potentials/lattice results within the estimated perturbative uncertainty.
\item
According to OPE, the leading non-perturbative
contribution to $E_{\rm tot}(r)$ in expansion in $r$ is ${\cal O}(r^2)$.
\item
Perturbative QCD prediciton of the singlet potential
$V_S(r)$ (leading Wilson coefficient in OPE) can be written as a
``Coulomb+linear'' potential.
Up to NNLL, the Coulomb+linear potential shows a converge towards lattice
results.
\end{itemize}
All the theoretical analyses conclude that, 
there is no freedom to add a linear potential of ${\cal O}(\LQ^2 r)$ 
to the perturbative prediction of the QCD potential at $r \simlt \LQ^{-1}$.
Therefore, if we are able to define the string tension from
the shape of the potential at  $r \simlt \LQ^{-1}$
(it has been empirically the case in phenomenological potential model approach),
the string tension may be within the reach of perturbative prediction.

\end{document}